\documentclass[reprint,aps,prb,superscriptaddress,showpacs]{revtex4-2}
\usepackage[utf8]{inputenc}
\usepackage{graphicx}
\usepackage{tikz}
\usepackage{dcolumn}
\usepackage{bm}
\usepackage{amsmath}
\usepackage{amsfonts}
\usepackage{amssymb}
\usepackage{hyperref}
\usepackage{pbox}
\usepackage{pinlabel}
\usepackage{ragged2e}
\usepackage[justification=RaggedRight]{caption}
\usepackage[singlelinecheck=off]{subcaption}
\usepackage{ulem}
\usepackage{dsfont}

\begin{document}
\title{Majorana fermions at self-generated interfaces}


\author{Nikola Prodanov}	
\affiliation{Dipartimento di Fisica, Universit\`a "Sapienza", Rome, Italy}	
\author{Sergio Ciuchi}%
\affiliation{Dipartimento di Scienze Fisiche e Chimiche, Università dell’Aquila, Coppito-L’Aquila, Italy}
\author{Sergio Caprara}
\affiliation{Dipartimento di Fisica, Universit\`a "Sapienza", Rome, Italy}


\date{\today}

\begin{abstract} 
The Kitaev model describing a one-dimensional topological superconducting chain is known to support two Majorana fermions localized at the system’s endpoints when the parameters are tuned to the topological phase. In this work, we investigate the possibility that Majorana fermions may also emerge away from the physical boundaries of the chain. To this purpose, we generalize the Kitaev model by incorporating a local coupling between the electronic density and a classical elastic (lattice) field. This electron–lattice interaction can induce phase separation between superconducting regions characterized by distinct topological invariants, thereby generating internal interfaces that host Majorana bound states. Under these conditions, a dilute gas of Majorana fermions can be realized in the bulk of the system.
\end{abstract}

\maketitle

\section{Introduction}{\label{Introduction}}
It is a remarkable fact that truly neutral fermions — whose particles
coincide with their antiparticles — first described by E. Majorana, can
be realized as elementary excitations in certain superconductors
endowed with specific topological properties \cite{Kitaev}. These excitations have zero energy (in the theromdynamic limit), and are often referred to as Majorana Zero Modes.
Although experimental
detections of Majorana Zero Modes have been reported \cite{deng2016majorana,YazdaniNature2017,SangjunScience2017,jackScience2019,shenPNAS2020,mannaPNAS2020}, in some cases subsequent controversy has led to retractions \cite{zhang2018retracted,zhang2021retraction,Gazibegovic2017retracted,Gazibegovic2017retractionnote}. From the theoretical point of view, models exhibiting Majorana fermions have proliferated over the last few decades, providing
guidance toward the practical realization of such superconducting devices.
Among the possible platforms, we mention here one-dimensional geometries,
such as nanowires \cite{deng2016majorana,YazdaniNature2017,mannaPNAS2020,SangjunScience2017} and edge states \cite{jackScience2019,shenPNAS2020}, vortex cores in superconductors \cite{sunnpj2017,sunScienceChina2017,zhuJApplPhys2021,HosurPRL2011,IoselevichPRL2011,ChiuPRB2011}, and Josephson vortices in heterostructures \cite{GrosfeldPNAS2011,PotterPRB2013}. For a recent review of possible platforms, see \cite{FlensbergNatRevMat2021}.

From a theoretical standpoint, intense research has been devoted to models exhibiting Majorana Zero Modes \cite{mazziotti2018majorana,VanHeckPRB2021,dvirMinimalKitaevNat2023,VanHeckPRB2026}, and numerous reviews have been published over the years \cite{AliceaRepProgPhys2012,SatoRepProgPhys2017,TanakaProgTheoExptPhys2024}.

The crucial property of Majorana fermions is that each one
represents half an electron. As a consequence, they appear in pairs in condensed
matter systems, are highly nonlocal objects in terms of the underlying
electron degrees of freedom, and possess braiding properties that make them
promising candidates for topological quantum computation \cite{microsoft2025interferometric,AasenMilestonesPRX2016}.
In one-dimensional geometries, Majorana fermions are typically localized
at the endpoints of the system \cite{Kitaev,CookPRB2011}. However, if the chain is
fragmented by disorder that is not strong enough to destroy the topological
properties, Majorana fermions can also emerge away from the endpoints.

Motivated by this observation, we investigate the possibility that
a low-density gas of Majorana fermions can form following the emergence of
self-generated interfaces, when the system is tuned to a regime in which
the ground state is phase-separated into two
superconducting phases with distinct topological properties.

The remainder of this paper is organized as follows. In Sec.\,\ref{model},
we introduce a model for a topological superconductor coupled to
a classical elastic field that contains all the necessary ingredients to
produce a phase-separated ground state, and we derive the equations that
self-consistently determine the deformation of the elastic
field induced by its coupling to the electron density. In
Sec.\,\ref{Results}, we present the phase diagram of the model and
discuss the occurrence of phase separation and the appearance of
Majorana fermions at the resulting self-generated interfaces. For
completeness, we also discuss the possible occurrence of charge-density
waves in the regime of very weak superconducting pairing. Concluding remarks
are presented in Sec.\,\ref{concl}. In Appendices
\ref{appa}, \ref{appb}, and \ref{appc}, we discuss, respectively,
the phase diagram in the special case of half-filling,
the calculation of the Green's function to determine the local
tunneling spectrum, and the occurrence of polarons as the final
fate of a disappearing interface.


\section{The Model}{\label{model}}
The simplest model to assess the occurrence of self-generated 
interfaces between two superconducting phases with different 
topological properties, is the Kitaev model equipped with a 
coupling between the local electron density and a classical 
elastic field, as described by the Hamiltonian
\begin{equation}\label{Hamiltonian with phonons}
\begin{split}
    H = & -t\sum_{l=0}^{N-1}(c_l^\dag c_{l+1} +c_{l+1}^\dag c_l)+ \Delta \sum_{l=0}^{N-1}(c_l^\dag c_{l+1}^\dag+c_{l+1}c_l) \\
    &-\mu\sum_{l=0}^{N-1}c_l^\dag c_l 
    +g\sum_{l=0}^{N-1} X_lc_l^\dag c_l+\sum_{l=0}^{N-1}\frac{1}{2}KX_l^2
\end{split}
\end{equation}
In this Hamiltonian, the first three terms on the right-hand side 
correspond to the standard Kitaev model: $l=0,...,N-1$ labels the 
sites of a linear chain, $t$ is a nearest-neighbor hopping, 
$\Delta$ is a nearest-neighbor pairing field, to enforce 
$p$-wave superconductivity in a system of spinless electrons, 
and $\mu$ is the chemical potential. If one assumes periodic 
boundary conditions, the site $N$ is to be identified with the 
site $0$. If, instead, open boundary conditions are used, the 
first two sums on the right-hand side of 
Eq.\,(\ref{Hamiltonian with phonons}) should stop at $N-2$. The 
classical elastic field at site $l$ is described by the variable
$X_l$. The term $gX_lc_l^\dag c_l$ denotes a linear 
interaction between the electron density at each site and the 
field variable, where $g>0$ quantifies the strength of this 
coupling. Finally, $\frac{1}{2}KX_l^2$ represents a harmonic 
(elastic) potential for the field, with $K>0$ determining 
the stiffness of this potential. In the following, we shall make 
use of the dimensionless coupling constant 
$\lambda\equiv\frac{g^2}{Kt}$. Since our model enjoys 
particle-hole symmetry, we can limit our analysis to electron 
densities $n\ge \frac{1}{2}$. The case $n=\frac{1}{2}$ corresponds
to half-filling (for spinless electrons, on average, one electron 
for two lattice sites). In this piece of work, we focus on the ground 
state properties, so we set the temperature $T=0$.

\subsection{Self-consistency equations}

The value of the local variable $X_l$ is self-consistently 
determined by the interplay between the elastic energy and 
the coupling to the local electron density. Indeed, minimization 
of the average energy $\langle H\rangle$ with respect to
$X_l$ yields $X_l=-\frac{g}{K}\langle c_l^\dag c_l \rangle$, i.e., 
apart from a trivial prefactor, the profile of the elastic field 
mirrors the electron density profile.

The translational invariance of the lattice can be spontaneously 
broken if $X_l$ assumes a value that is explicitly site-dependent. 
For instance, it is well known that the model in 
Eq.\,(\ref{Hamiltonian with phonons}) is prone to a Peierls 
instability when $\Delta=0$, for any finite $g$. At weak coupling, the 
system hosts a charge-density wave with wave vector $2k_F$, where 
$k_F$ is the electron density-dependent Fermi wave vector of the 
system at $g=0$. A particular case is the dimerization at 
half-filling, when $\mu=0$ and $k_F=\frac{\pi}{2a}$, where $a$ is 
the lattice spacing. Then, e.g.,
$\langle c_l^\dag c_l \rangle=\frac{1}{2}[(-1)^l+1]$, and $X_l$ 
oscillates accordingly. However, for a fixed $g$, a sufficiently 
large $\Delta$ eventually suppresses the tendency to charge-density 
waves, and enforces a uniform electron density.

The model in Eq.\,(\ref{Hamiltonian with phonons}) can be studied 
in real space, where the task of diagonalization of the Hamiltonian 
and self-consistent determination of the local value of the 
classical elastic field, and of the value of the chemical potential 
to fix a given electron density $n$ (for a total of $N+1$ self-
consistency equations), can be accomplished numerically, even on 
quite large systems ($N\approx 10^3$). These numerical 
solutions indeed exhibit all the possible phases of the
model. 

Nonetheless, 
many solutions in real space can 
be readily interpreted as phase-separated solutions, where two 
phases coexists, with one or more self-generated interfaces 
between them. Furthermore, for reasonable values of the
parameters, the Peierls-like charge-density wave phase is limited 
to a narrow region of the phase diagram (see below, Sec.\,\ref{cdw}). 
So, a good understanding of the physics of the model in 
Eq.\,(\ref{Hamiltonian with phonons}) can be gained studying 
homogeneous phases, with $X_l=X_0=-\frac{g}{K}n$, where $n$ is 
the average electron density, and dimerized phases, 
with $X_l=X_0+(-1)^l X$, where a nonzero $X$ enforces the
dimerization of the lattice. The homogeneous phase does not break 
the lattice periodicity, while the dimerized phase simply doubles 
it, so both phases can be studied in reciprocal space, with 
periodic boundary conditions, making the treatment numerically 
much simpler. Indeed, in this case, the parameters to 
be self-consistently determined are only three, $X_0,X,\mu$, 
with $X=0$ in the homogeneous phase, and the eigenvalues of 
the Hamiltonian, for a generic set of parameters $X_0,X,\mu$, can 
be determined analytically. 

In the presence of a dimerized elastic field $X_l=X_0+(-1)^l X$,
Eq.\,(\ref{Hamiltonian with phonons}) can be diagonalized
in reciprocal space, after the Brillouin zone of the original 
lattice has been folded form $[-\frac{\pi}{a},\frac{\pi}{a}]$ 
to $[-\frac{\pi}{2a},\frac{\pi}{2a}]\equiv D$. The spectrum can 
be determined analytically by means of the Nambu-Gor'kov 
formalism, yielding
\begin{equation}
    E_k^{\eta,\nu}=\eta\sqrt{\varepsilon_k^2+\Delta_k^2+
    \tfrac{1}{4}\tilde\mu^2
    +\tfrac{1}{4}g^2X^2+\nu R_k}\, ,
    \label{eq-ek}
\end{equation}
where $\eta=\pm$ distinguishes the two Nambu eigenvalues, $\nu=\pm$,
$\varepsilon_k\equiv -t\cos(ka)$. $\Delta_k\equiv \Delta\sin(ka)$, 
$\tilde{\mu}\equiv\mu-gX_0$, $R_k\equiv\sqrt{(\varepsilon_k\tilde{\mu})^2
+(gX\Delta_k)^2+\frac{1}{4}(gX\tilde{\mu})^2}$.
The grand-canonical 
thermodynamic potential per lattice site at a temperature $T=0$ is
simply $\omega=\frac{1}{N}\langle H\rangle$. The requirement that 
$n=-\frac{\partial\omega}{\partial\mu}$, and minimization of $\omega$ 
with respect to $X_0$ and $X$, yield three 
coupled self-consistency equations:
\begin{equation}\label{equations of minimizations}
\left\{
    \begin{split}
        &n=\frac{1}{2}+\frac{\Tilde{\mu}}{2N}\sum_{\nu,k \in D}\frac{1}{E_k^{+,\nu}}\left[\frac{1}{2} + \nu\frac{\varepsilon_k^2+\Big(\frac{gX}{2}\Big)^2}{R_k} \right],\\
        &X_0=-\frac{g}{K}n,\\
        &X=\frac{g^2X}{2NK}\sum_{\nu,k \in D}\frac{1}{E_k^{+,\nu}}\left[\frac{1}{2} + \nu \frac{\Delta_k^2+\Big(\frac{\tilde{\mu}}{2}\Big)^2}{R_k} \right].\\
    \end{split}\right.
\end{equation}
Inserting the second equation in\,(\ref{equations of minimizations})
in the definition of $\tilde\mu$, we can explicitly write 
$\tilde\mu$ in terms of $\mu$ and $n$, as 
\begin{equation}\label{eq:mu symmetric mu relation}
    \tilde{\mu}=\mu+\frac{g^2}{K} n=\mu+\lambda t n.
\end{equation}

Given the self-consistency equations
(\ref{equations of minimizations}), we can find numerical solutions 
for $\tilde{\mu}$ and $X$, the solution for $X_0$ being explicit. 
We notice, on passing, that by setting $\tilde{\mu}=0$ the first 
equation indicates that the system is at half filling.

Once the self-consistency equations are solved, the free energy per 
lattice site, $f=\omega+\mu n$, reads
\begin{equation}
    f=-\frac{1}{N}\sum_{\nu, k \in D}E_k^{+,\nu}+\frac{1}{2}KX^2-\Tilde{\mu}\left( n-\frac{1}{2}\right)-\frac{g^2 n^2}{2K}.
\end{equation}

It is worth noticing that our model becomes formally equivalent to the
Kitaev model in the homogeneous phase ($X=0$), provided the chemical
potential is identified with $\tilde\mu$, see 
Eq.\,(\ref{eq:mu symmetric mu relation}). Then, the boundary 
between the topological and non-topological phase is marked 
by the topological criterion of the Kitaev model, 
$|\tilde{\mu}|=2t$. The first self-consistency equation 
in\,(\ref{equations of minimizations}), in the absence of 
dimerization ($X=0$), yields the density as a function of 
$\tilde\mu$, $n=n(\tilde{\mu},\Delta)$. Setting $\tilde{\mu}=2t$ 
(we are only considering the case $n\ge\frac{1}{2}$), eliminates the 
dependence on $\lambda$, showing that the boundary between the 
topological and non-topological phase is reached at a density 
$n_{\mathrm{topo}}=n(2t,\Delta)$ that is independent of 
$\lambda$. Indeed,
\begin{equation}\label{eq:analytics expression pd kitaev}
    n_{\mathrm{topo}}=\frac{1}{2}+\frac{1}{2N}\sum_k \frac{1}{\sqrt{1+ \left(\frac{\Delta}{t}\right)^2\cot^2\left(\frac{k}{2}\right)}},
\end{equation}
where the sum over $k$ is extended to the entire Brillouin zone of 
the original (undimerized) lattice. Then, from 
Eq.\,(\ref{eq:mu symmetric mu relation}), we can determine 
the value of the chemical potential in correspondence 
of which the change of topological properties occurs,
\begin{equation}
    \mu_{\mathrm{topo}}=(2-\lambda \,n_{\mathrm{topo}})\,t,
\end{equation}
which does also depend on $\lambda$.

\section{Results}{\label{Results}}
This section is devoted to a discussion of the properties of our 
model, Eq.\,(\ref{Hamiltonian with phonons}), as obtained by the
numerical solution of the self-consistency equations in real space,
and the insight gained from the analysis of the homogeneous and dimerized solutions in reciprocal space.

\begin{figure}[h!]
\includegraphics[width=9cm]{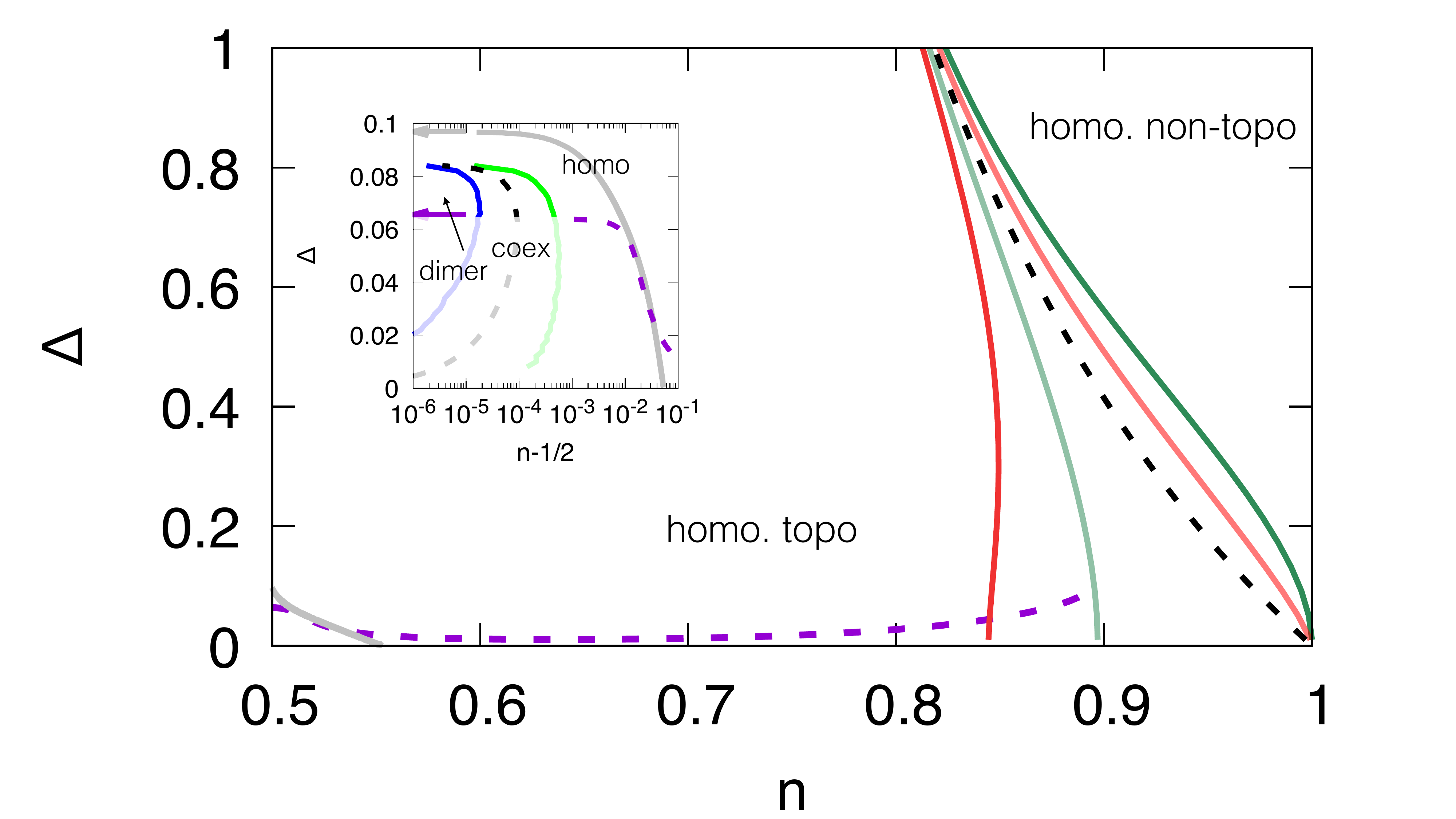}
\caption{a) Phase diagram for $\lambda=2$. Red and green lines results 
from Maxwell construction respectively separating non-topological and 
topological homogeneous phases. Black dashed line marks the occurrence 
of topological criterion when $\lambda=0.0$ and it is the locus of the 
first order phase transition when $\lambda>0$. Light green (light red) 
lines are spinodal lines of the red (green) topological (non-
topological) homogeneous phases. Purple dashed line marks the 
instability line of the homogeneous topological phase towards a CDW 
(see Sec.\,\ref{cdw}); this line meets the spinodal line of the 
homogeneous topological phase at $n\approx 0.9$ and 
$\Delta\approx 0.1$, and merges with it.  Grey line is the spinodal line of the dimer above which dimer is no longer a metastable minimum of the free energy.
The inset shows an enlarged view of the region near $n=\tfrac{1}{2}$, 
where a first-order phase transition between a dimerized 
non-topological phase (dimer) and a homogeneous topological 
phase occurs (black dashed line). Around this line, a coexistence 
region can be determined via Maxwell construction (marked here by bue and green lines), leading to phase 
separation. Below the purple line the lines are shaded indicating the possibility of Peierls CDW ground state.}
\label{fig:pd}
\end{figure}

The most striking result of our analysis is that, at $\lambda>0$, 
the line that separates the topological (at lower $n$) and 
non-topological (at larger $n$) homogeneous phases is embedded in 
a coexistence region that widens with increasing $\lambda$, for 
fixed $\Delta$, and shrinks with increasing $\Delta$, at fixed 
$\lambda$, see Fig.\,\ref{fig:pd}.

When $\Delta$ is small enough, the homogeneous phase competes with 
a Peierls-like charge-density wave (see Sec.\,\ref{cdw}), and 
in particular with a dimerized phase ($X\neq 0$), near half filling 
(see inset of Fig.\,\ref{fig:pd}). Here, a coexistence region 
occurs between a non-topological dimerized phase and a topological 
homogeneous superconducting phase. More detail about the physics 
of our model at half filling is given in Appendix\,\ref{Half-filling}. 
Hereafter, we focus on the more interesting region of coexistence 
between two homogeneous phases with different topological 
properties, that is found whenever $\Delta$ is sufficiently large.
The origin and properties of this phase-separation region are 
discussed in the next section.

\subsection{Phase separation}

\begin{figure}[h!]
  \includegraphics[width=8cm]{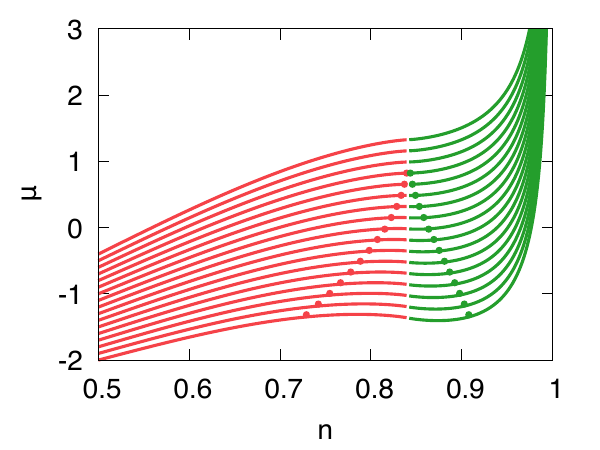}
    \caption{The chemical potential versus the density of electrons 
    $n$ at $\Delta=0.8t$ and for several equally spaced values of 
    $\lambda$, from $\lambda=1.0$ (top curve) to $\lambda=4.0$ (bottom 
    curve). On the green branches (non-topo) the topological criterion 
    of the Kitaev model is not satisfied, while it is met on the red 
    branches (topo). Small dots marks the cohexistence region by Maxwell's construction. The border between non-topo and topo branches do not depend on $\lambda$ as predicted by Eq. (\ref{eq:analytics expression pd kitaev}).}
    \label{fig: mu vs n: topo vs nontopo homogeneous}
\end{figure}

The coexistence region between two homogeneous phases in the 
phase diagram of Fig.\,\ref{fig:pd} is a consequence of the fact 
that at $\lambda>0$, while $n$ is a monotonic function of 
$\tilde\mu$, for $X=0$ (homogeneous phase), 
$\mu=\tilde\mu-\lambda t n$ becomes non-monotonic in a certain 
electron density window (spinodal region), for a 
sufficiently large $\lambda$, see 
Fig.\,\ref{fig: mu vs n: topo vs nontopo homogeneous}. This 
non-monotonic behavior indicates that the system is not stable in 
a single phase, rather it separates into two phases, whose 
densities. $n_1$ and $n_2$, are determined by the 
Maxwell construction, cutting the non monotonic 
portion of the $\mu$ vs. $n$ curves with a horizontal line, in such 
a way that the two lobes have equal areas.

The phase separation region encompasses the spinodal region. The 
two coexisting phases have each a fixed density (determined by 
the Maxwell construction), so that in one phase 
the density is $n_1<n_{\mathrm{topo}}$ (topological phase)
and in the other phase 
the density is $n_2>n_{\mathrm{topo}}$ (non-topological phase).
With increasing $n$, in the window $n_1\le n\le n_2$, each phase 
stays at a fixed density, while the average density $n$ increases 
by increasing the fraction $x$ of the phase at higher density, 
$n=(1-x)n_1+xn_2$, with $0\le x\le 1$.

\subsection{Majorana fermions at interfaces}

\begin{figure}[h!]
  \includegraphics[width=9cm]{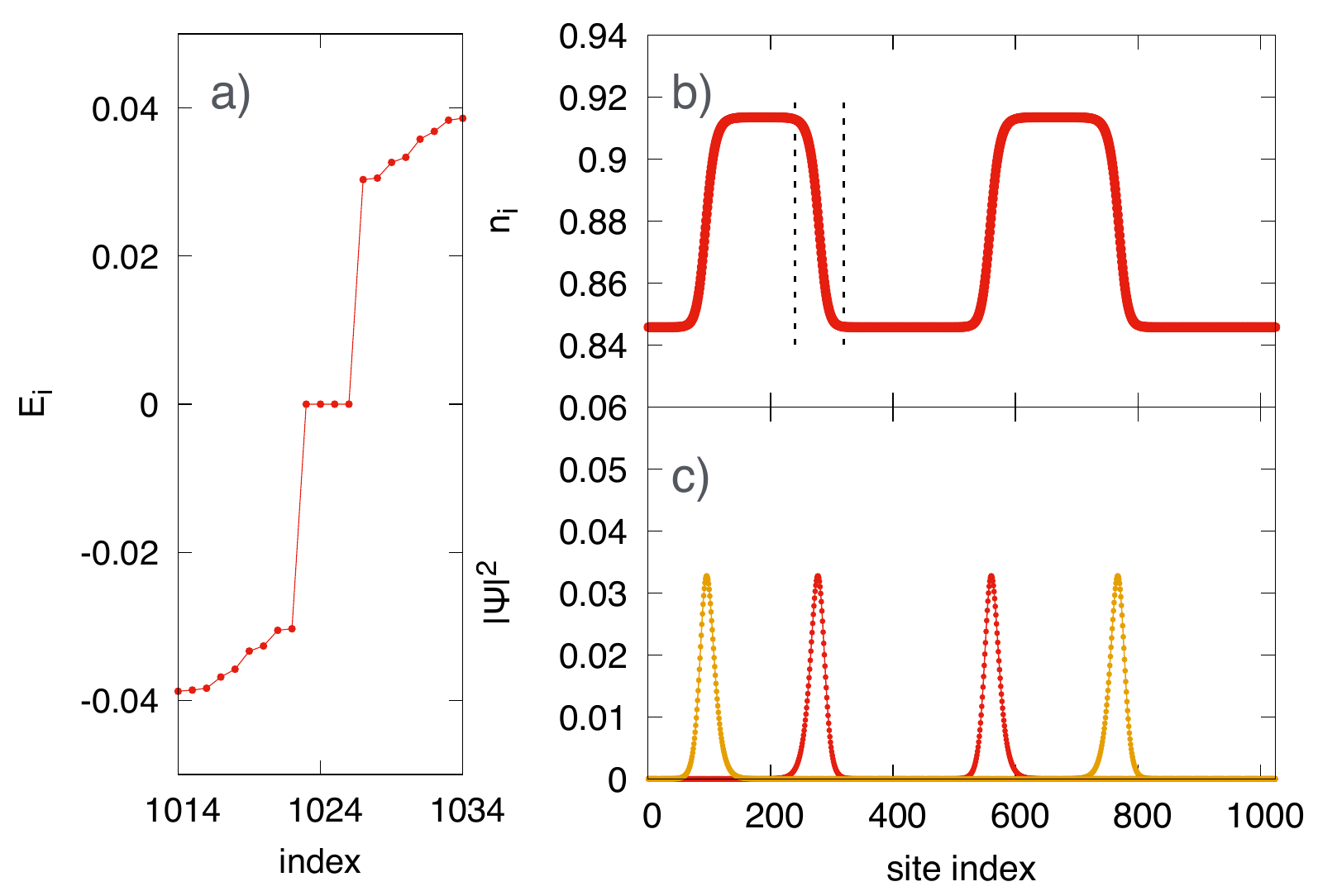}
    \caption{Properties of a numerical solution obtained in the 
    coexistence region of Fig.\,\ref{fig:pd}, for 
    $\lambda=2,\,\Delta=0.5$, with periodic boundary conditions. a) 
    Eigenvalues showing 4 zero modes. b) Charge modulation in the 
    $N=1024$ chain. The interface located between two dashed lines is 
    sampled in Fig.\,\ref{fig:GreenFunction}. c) The modulus squared 
    of the Majorana fermion wave function, as a function of the site 
    index (colors mark the pairwise related zero-modes).}
    \label{fig: NTT charge}
\end{figure}

Now we can discuss the most important consequence of the coexistence
of two homogeneous phases with different topological properties within 
our model. Whenever the Kitaev model is tuned in the topological 
phase, two Majorana fermions appear at the ends of the chain, if 
open boundary conditions are adopted. If, however, one adopts 
periodic boundary conditions, the two Majorana fermions re-bind 
into an electron and disappear, even if the topological criterion 
is still obeyed. Within our model, instead, if we look for 
solutions of the self-consistency equations in real space, 
within the coexistence region, starting from random 
initial configuration, even if we adopt periodic 
boundary conditions, we find a certain number of self-generated 
interfaces hosting Majorana fermions. 
For instance, in Fig.\,\ref{fig: NTT charge}\,a) we display 
the spectrum corresponding to a numerical solution of the 
self-consistency equations in real space, exhibiting four 
zero-energy modes. The 
electron density profile in Fig.\,\ref{fig: NTT charge}\,b)
shows that four interfaces have been generated, as a result of
the formation of two droplets of the minority phase embedded in 
the majority phase. 

Each interface has a finite width, across which 
the electron density profile smoothly interpolates between $n_1$ 
and $n_2$, and hosts a Majorana fermion, whose maximum is found 
more or less in the middle of the interface region,
see Fig.\,\ref{fig: NTT charge}\,c). The Majorana fermions are 
pairwise related, each pair occurring at the endpoints of a
fragmented domain of the topological phase.

\begin{figure}[h!]
\includegraphics[width=9cm]{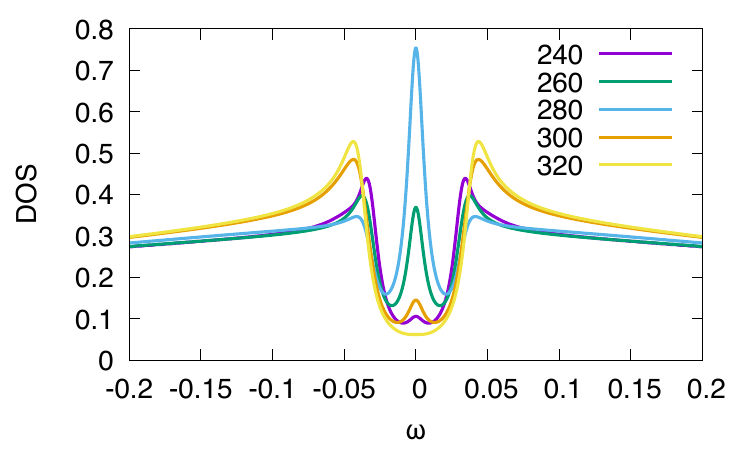}
\caption{Local spectral density across the interface shown in 
Fig.\,\ref{fig: NTT charge}\,b) (dashed lines), for 
$\lambda=2,\Delta=0.5$ and a broadening $7\times 10^{-3}$. 
The numerical labels of the various lines identify the site along 
the chain where the corresponding spectral density is found.}
\label{fig:GreenFunction}
\end{figure}

We conclude this section with a brief discussion on the detection 
of Majorana fermions. A direct calculation of the 
local spectral density across the interface delimited by two 
dashed lines in Fig.\,\ref{fig: NTT charge}\,b) shows that 
the Majorana fermion appears as a zero-bias peak, 
see Fig.\,\ref{fig:GreenFunction}. Details about the
calculation of the Green's function, that allows to determine
the local tunneling spectra, are given in Appendix\,\ref{appb}.

The above result suggests that Majorana fermions might be detected in 
local tunneling measurements. Unfortunately, many other mechanisms 
may provide spectral weight at zero bias, due, e.g, to
trivial Andreev bound states \cite{LiuAndreevBoundStatePRB2017}, interface disorder \cite{DasSarmaNatPhys2023}, or 
non-topological low-energy states \cite{LaubscherCDGMPRB2025} generated by strong magnetic 
fields required to reach the effective spinless regime in realistic condensed-matter systems. 
Furthermore, polarons, that do appear at non-zero energy 
whenever two droplets of the non-topological homogeneous phase coalesce in the
phase-separation region
(see Appendix\,\ref{appc}), may have spectral weight at zero energy 
in the presence of a modest broadening.
Therefore, it would be of great importance to devise an experimental 
procedure that gives an unambiguous and definitive answer about 
the existence of bona-fide Majorana fermions. 

\subsection{Peierls instability and charge-density waves}
\label{cdw}

\begin{figure}[h!]
  \includegraphics[width=9cm]{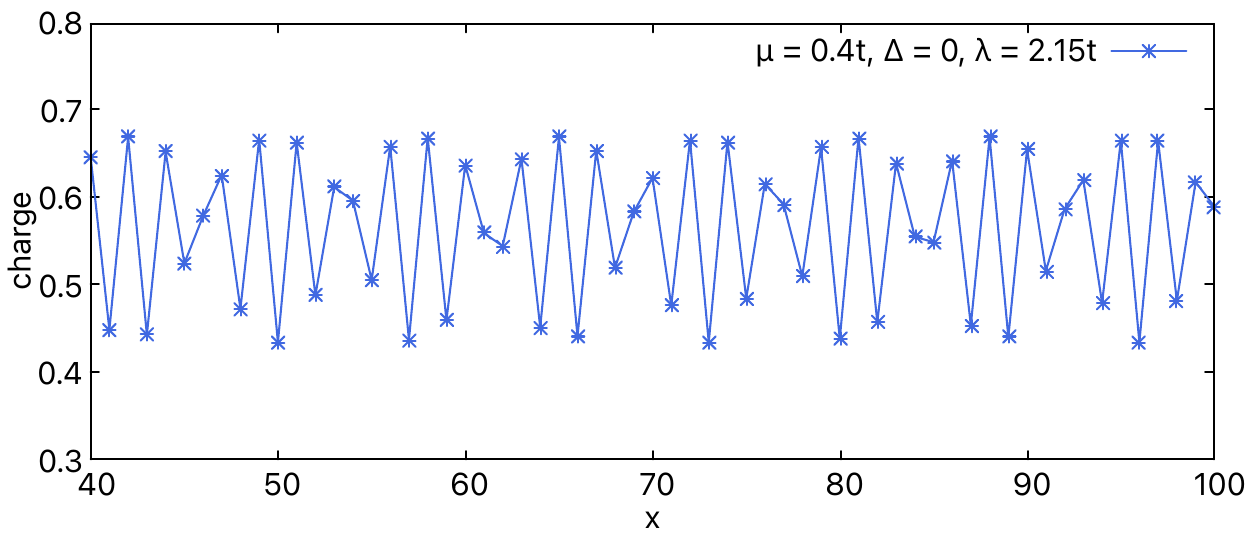}
    \caption{A charge-density wave generated by the Peierls instability. Here, $t=1$, $\mu=0.4$ (i.e., $n\approx 0.55$), $\Delta=0$, $\lambda=2.15$, and $N=200$. The charge is expressed in units of the electron charge $e$. The density profile results from computational minimization of the free energy.}
    \label{fig: trimmer}
\end{figure}

The model of Eq.\,(\ref{Hamiltonian with phonons}) is known to 
undergo a Peierls instability towards a charge-density wave at 
any $g>0$, when $\Delta=0$. This is due to a 
divergence of the static charge susceptibility $\chi_c(q)$ for a wave 
vector $q=2k_F$, where $k_F$ is the Fermi wave vector. An example of a 
charge-density wave with wavelength 
$\Lambda_{\mathrm{CDW}}\approx 25\,a$ is shown in 
Fig.\,\ref{fig: trimmer}.

A finite $\Delta$ rapidly suppresses this tendency, because of the
opening of the gap around at the Fermi surface, as confirmed by the 
calculation of the static charge susceptibility in the 
homogeneous phase
\begin{equation}
\chi_c(q)=\frac{1}{2N}\sum_k \frac{E_+ E_- - \xi_+ \xi_- +
\Delta_+\Delta_-}{E_+ E_- (E_+ + E_-)}
\end{equation}
where $E_\pm\equiv E_{k\pm q/2}$, $\xi_\pm\equiv \xi_{k\pm q/2}$,
and $\Delta_\pm\equiv \Delta_{k\pm q/2}$, with 
$\xi_k\equiv -t\cos(ka)-\mu$, $\Delta_k\equiv\Delta\sin(ka)$, and
$E_k\equiv\sqrt{\xi_k^2+\Delta_k^2}$.
The homogeneous phase becomes unstable towards a charge-density 
waves only for small values of $\Delta$ and not too large $n$. 
In Fig.\,\ref{fig:pd}, the Peierls instability line for the 
homogeneous phase is marked by a purple dashed line. 
Remarkably, at large density (or large $\Delta$), the 
instability moves from finite $q$ to $q=0$, which marks a phase separation. Indeed, the 
Peierls instability line joins the spinodal line in the region of 
coexistence of two homogeneous phases with different topological 
properties. Near half filling, instead, the charge-density wave 
is commensurate with the lattice and promotes a dimerization 
($q=\frac{\pi}{a}$). Since the dimerized phase is confined to the 
region of our phase diagram where $\Delta$ is small and $n\approx \frac{1}{2}$ ,
we shall not discuss it any further. The reader interested in 
more detail may refer to Appendix\,\ref{Half-filling}.

\section{Discussion and conclusions}
\label{concl}
Motivated by the fact that Majorana fermions are typically found
at the endpoints of one-dimensional models for topological
superconductors, but can be displaced from the endpoints
when the system is fragmented by a disordered potential that is not
strong enough to suppress the topological properties,
we investigated the possible occurrence of
a low-density gas of Majorana fermions in a model for a topological superconductor
that can be tuned to a phase-separated ground state.
When the two coexisting phases have distinct topological properties,
Majorana fermions are expected to appear at
self-generated interfaces.

The simplest realization of this physical scenario is
achieved by the one-dimensional Kitaev model for spinless
fermions, in which the electron density is locally coupled to
a classical elastic field. The most significant result of our analysis
is that a substantial portion of the phase diagram hosts a coexistence
region, in which the ground state is phase-separated into a
lower-density topological phase and a higher-density nontopological
phase. The numerical solution shows that interfaces separating
droplets of the minority phase from the majority phase are
self-generated, and that Majorana fermions indeed appear at these
interfaces. Interestingly, we also find that whenever two droplets
of the nontopological phase coalesce within the coexistence region,
a single hole becomes trapped at their interface (or an electron,
at filling $n<\tfrac{1}{2}$, by particle-hole symmetry), and a
polaron is formed as a bound state within the superconducting gap.
Unlike the Majorana fermions, this polaron state is not pinned at
zero energy. As soon as the boundary of the coexistence region is
crossed, these polarons disappear. Further details on the properties
of these polarons are provided in Appendix\,\ref{appc}.

We now discuss the robustness of our results with respect to
the relaxation of some of our approximations. Concerning the
stability of the phase-separated state against quantum and thermal
fluctuations: at zero temperature, the phase-separated state is
expected to be stable against quantum fluctuations of the phonon
field, and therefore the proposed mechanism should remain operative
even in the presence of phonons with finite frequency. This is,
however, no longer the case at finite temperature. In one dimension
and in the absence of externally imposed charge segregation, the
proliferation of thermal defects in the charge-ordered phase
effectively reduces the spatial separation between the boundary
Majorana Zero Modes. Nevertheless, such defect proliferation can be
suppressed by an applied external gate potential, by the coupling to the proximizing bulk superconductor, or may occur
only on rather long time scales, thereby preserving the
spatial separation and stabilizing the Majorana Zero Modes over
a well-defined temporal window. 

A further relevant consideration concerns the assumption of spinless
fermions in the Kitaev model. This condition can in principle be
physically realized through spin-orbit coupling in the presence of
an external magnetic field.

Finally a possible extension of the present work would be the study of the interaction with elastic degree of freedom coupled to the hopping {\it a la} Su-Schrieffer-Heeger. In this case the dimerization will be bond instead of site centered and the local density is expected to be constant thereby preventing the competition with the pairing field.

\appendix
\section{Phase Diagram at Half-filling}{\label{Half-filling}}
\label{appa}

\subsection{Dimerization}
\begin{figure}[h]
   \includegraphics[width=9cm]{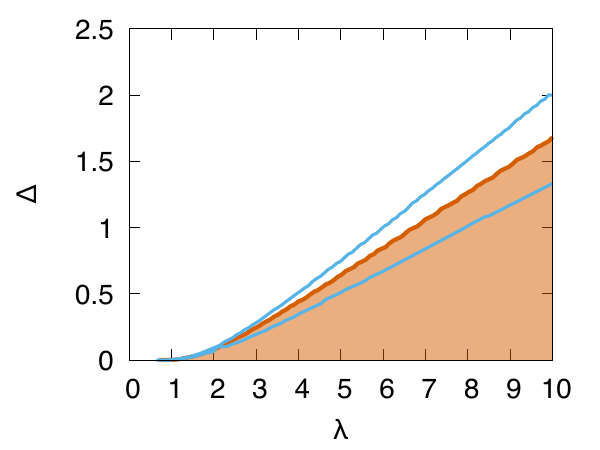}
 \caption{Phase diagram at half-filling. The first-order
 transition line (red) between the homogeneous topological phase (white region)
 and the dimerized non-topological phase (reddish region) is located inside
 a coexistence region, delimited by two spinodal lines (blue curves).}
\label{fig:phasediag-half-filling}
 \end{figure}

At half filling ($n=\tfrac{1}{2}$), Eq.\,(\ref{eq-ek}) 
simplifies, because particle-hole symmetry imposes 
$\tilde\mu=0$, hence
\begin{equation}
    E_k^{\eta,\nu}=\eta\sqrt{\varepsilon_k^2+\left(\Delta_k+
    \tfrac{\nu}{2}gX\right)^2}\, .
\end{equation}
The first of the three Eq.s\,(\ref{equations of minimizations})
is solved trivially when $\tilde\mu=0$. The second equation 
yields $X_0=-\frac{g}{2K}$, so one is left with one equation to
solve, namely
\begin{equation}
       X=\frac{g}{2NK}\sum_{\nu,k \in D}
       \frac{\left(\tfrac{1}{2}gX + \nu \Delta_k 
       \right)}{E_k^{+,\nu}}.
\end{equation}
The free energy per lattice site simplifies to
\begin{equation}
    f=-\frac{1}{N}\sum_{\nu, k \in D}E_k^{+,\nu}+\frac{1}{2}KX^2-\frac{g^2}{8K}.
\end{equation}
The homogeneous phase, with $X=0$, is topological, while the dimerized phase, with $X\neq 0$, is non-topological. The two phases are separated by a first-order phase transition (see 
Fig.\,\ref{fig:phasediag-half-filling}), which is the line along which the free energies of the two phases are equal. A region where the two
phases coexist as local minima of the free energy, delimited by two spinodal lines, is found around
the transition line. Looking at Fig.\,\ref{fig:phasediag-half-filling}), the upper spinodal line is the line above which the dimerized phase ceases to be a local minimum of the free energy, the lower spinodal line is the line below which the homogeneous phase ceases to be a local minimum of the free energy
The analysis carried out in Sec.\,\ref{Results} at a generic filling $n$, shows that this coexistence region originates a phase separation region between the homogeneous and the dimer phase (see the inset of
Fig.\,\ref{fig:pd}).

 \subsection{Zero modes at half-filling}
 \begin{figure}[h!]
  \includegraphics[width=9cm]{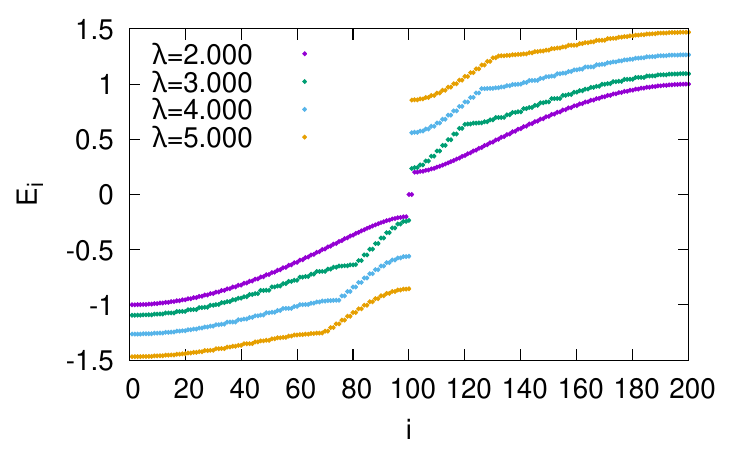}
    \caption{Eigenvalues of the Kitaev chain at half-filling for $N=100$, $t=1$, $\Delta=0.2$ and different couplings $\lambda$.}
    \label{fig: HF real space spectrum}
\end{figure}

In Fig.\,\ref{fig: HF real space spectrum}, we show the energy spectrum at the global minima of the free energy and half filling, for $N = 100$, $t=1$,
$\lambda \in (2,\,3,\,4,\,5)$, and $\Delta = 0.2$. Here, 
$i = 0, 1, \ldots, 2N-1$ labels the energy eigenvalues $E_i$ 
in increasing order. For $\Delta=0.2\,t$, the critical value of $\lambda$ at which the first-order phase
transition occurs is $\lambda_c = 2.79\,t$. For any 
$\lambda < \lambda_c$, the global minimum of the free energy
corresponds to the homogeneous phase, the topological 
criterion is met, the spectrum is independent
of $\lambda$ and features two zero modes. Conversely, for $\lambda > \lambda_c$, the global minimum of the free energy corresponds to 
the non-topological dimerized 
phase, the spectrum depends on $\lambda$ and zero modes
are absent.

\section{The Green's function}
\label{appb}

Here, we resume the basic step to derive the local Green's function in the presence of an external pairing field, as it appears in the Hamiltonian of the main text. The procedure can be considered a generalization of the one described in 
Ref.\,\cite{Viswanath-Muller}.

We basically consider the inversion of a block matrix in the site representation:
\begin{eqnarray}
G^{-1}=\left( \begin{array}{cccc}
a_0 &  b_0 & 0 &...\\
 b^\dagger_0 & a_1 & b_1 & ...\\
...&...&...&...\\
...&0& b^\dagger_{N-1} & a_N \end{array} \right),
\end{eqnarray}
where each block named $a_i,b_i$ is a $2\times 2$ matrix which in our case reads
\begin{eqnarray}
a_i = (\omega-\epsilon_i) \mathds{1},\\
b_i = t_{i,i+1}\mathds{1}-\Delta_{i,i+1} \sigma_x,
\label{eq:InvG}
\end{eqnarray}
$i$ is a site index of the chain, $\sigma_x$ is a Pauli matrix, and one might also consider complex valued and site dependent hoppings $t$ and couplings $\Delta$, and site dependent energies $\epsilon$.

To write a recursion formula for the diagonal elements of the inverse $G$ is better to isolate a generic site of the chain and regroup the blocks of the matrix $G^{-1}$ as the following
\begin{eqnarray}
G^{-1}=\left( \begin{array}{ccc}
A_{k-1} &  B_{k-1} & 0 \\
B^\dagger_{k-1} & a_k & B_k \\
0& B^\dagger_{k} & A_{k+1} \end{array} \right).
\label{eq:blockInvG}
\end{eqnarray}

Using the same block convention we define the Green's 
function matrix 
as
\begin{eqnarray}
G=\left( \begin{array}{ccc}
G_{k-1} &  F_{k-1} & U_{k-1} \\
F^\dagger_{k-1} & g_k & F_k \\
U^\dagger_{k}&F^\dagger_k & G_{k+1} \end{array} \right),
\label{eq:blockG}
\end{eqnarray}
where $g_k$ is the Green's function ($2\times 2$) matrix that we want to calculate.

Exploiting the relation $G^{-1}G=\mathds{1}$, we have
\begin{equation}
B^\dagger_{k-1}F_{k-1}+a_k g_k + B_k F^\dagger_{k}=\mathds{1}.
\label{eq:eqdiag}
\end{equation}

From the upper diagonal matrix block we get
\begin{equation}
A_{k-1}F_{k-1}+B_{k-1} g_k =0.
\label{eq:equpperdiag}
\end{equation}

From the lower diagonal matrix block we have
\begin{equation}
B^\dagger_{k}g_{k}+A_{k+1} F^\dagger_{k}=0.
\label{eq:eqlowerdiag}
\end{equation}

Using Eq.s\,(\ref{eq:equpperdiag},\ref{eq:eqlowerdiag}) to eliminate $F$ in Eq.\,(\ref{eq:eqdiag}), we get the Dyson equation for the Green's function matrix
\begin{equation}
g_k = a^{-1}_k + a^{-1}_k \left [ \Sigma^{(R)}_k+ \Sigma^{(L)}_k\right ] g_k,
\label{eq:Dyson}
\end{equation}
where
\begin{eqnarray}
\Sigma^{(R)}_k &=& B_k A^{-1}_{k+1} B^\dagger_k,\label{eq:defSR}\\
\Sigma^{(L)}_k &=& B^\dagger_{k-1} A^{-1}_{k-1} B_{k-1}.\label{eq:defSL}
\end{eqnarray}
We note here from Eqs.\,(\ref{eq:InvG},\ref{eq:blockInvG}) that the rectangular matrix $B_k$ has only one $2\times 2$ block different from zero therefore the previous two equations give $2\times 2$ matrices. The elements $A^{-1}_{k+1}$ and $A^{-1}_{k+1}$ are the Green's function matrices evaluated respectively at the right and the left edge of the cut open when the site $k$ has been removed.

By considering the actual structure of the block tridiagonal inverse Green's function matrix, Eq.\,(\ref{eq:InvG}), we can construct a recursion from the left and right self-energies appearing in Eq.s\,(\ref{eq:defSR},\ref{eq:defSL}). Indeed the evaluation of the edge Green's function when the site $k$ has been removed is simple since it involves respectively only the left and the right self-energies
\begin{eqnarray}
\Sigma^{(R)}_k &=& b_k \left [ a_k - \Sigma^{(R)}_{k+1}\right ]^{-1} b^\dagger_k, \label{eq:recSR}\\
\Sigma^{(L)}_k &=& b^\dagger_{k-1} \left [ a_{k-1} - \Sigma^{(L)}_{k-1}\right ]^{-1} b_{k-1}.\label{eq:recSL}
\end{eqnarray}

The solution of the Dyson equation Eq. (\ref{eq:Dyson}) 
\begin{equation}
g_k = \left [ a^{-1}_k - \Sigma^{(R)}_k- \Sigma^{(L)}_k\right ]^{-1},
\label{eq:LocalG}
\end{equation}
together with the iterative recursions in
Eq.s\,(\ref{eq:recSR},\ref{eq:recSL}), gives the local Green's function matrices in both normal and anomalous matrix elements.







\begin{figure}
    \centering
    \includegraphics[width=0.7\linewidth]{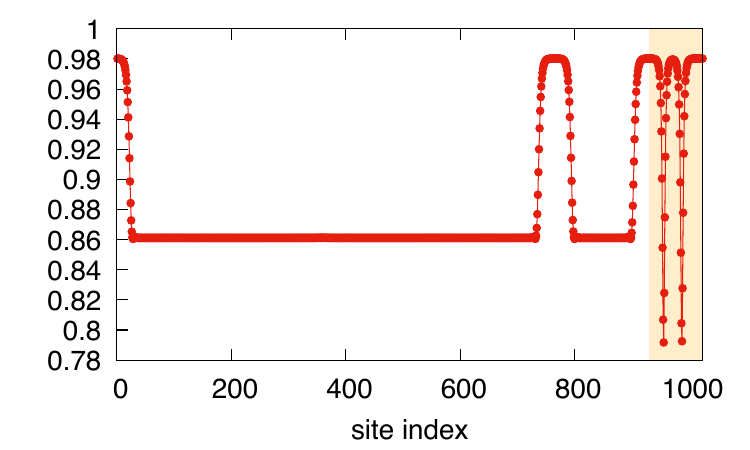}\\
    \includegraphics[width=0.7\linewidth]{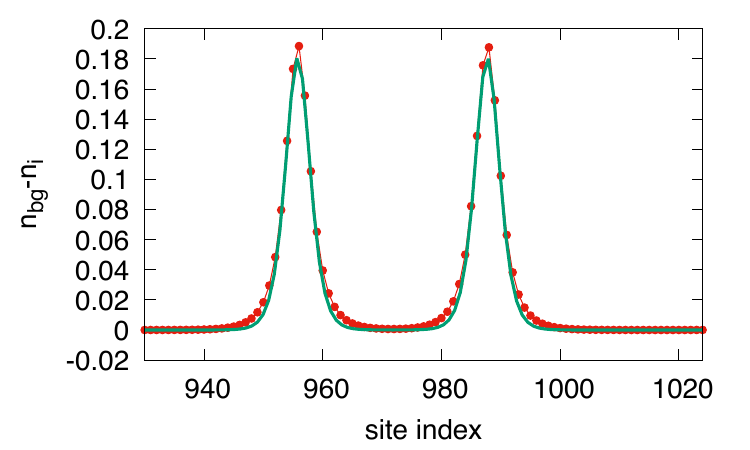}\\
    \includegraphics[width=0.7\linewidth]{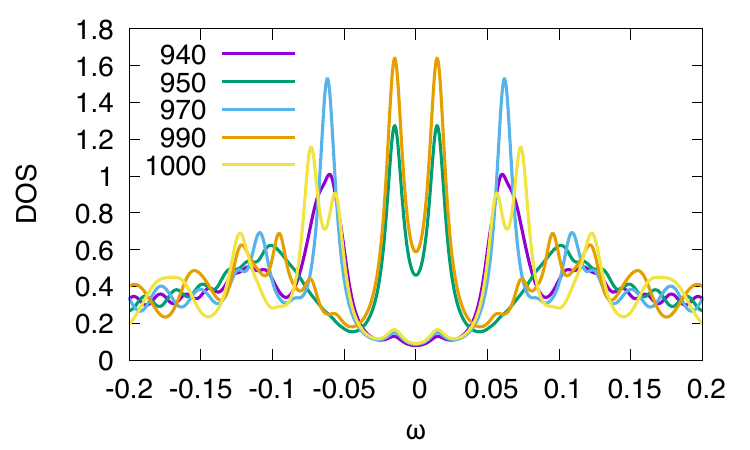}\\
    \caption{Upper panel: density profile obtained when $t=1$, $\mu=1.9,\,\lambda=2,\,\Delta=0.2$, near the spinodal line of the non-topological homogeneous phase. The orange area marks the occurrence of two polarons at the interfaces of three coalescing droplets of the non-topological (high-density) phase. Middle panel: polaronic density, obtained as the difference between the nominal density of the minority phase (background) and the actual density profile ($n_\mathrm{bg}-n_i$), fitted with the strong-coupling one dimensional estimate (see text).
    Lower panel: Local spectral density using a broadening $7 \times 10^{-3}$. The numerical labels of the 
    various lines identify the site along the chain where the 
    corresponding spectral density is found.}
    \label{fig:polaron}
\end{figure}

\section{Polarons}
\label{appc}
As we discussed in Sec.\,\ref{concl}, the numerical solution 
of the self-consistency equations of our model shows that
whenever two droplets of the topological phase coalesce in the coexistence region, a single hole is eventually trapped at the 
vanishing interface between them, and a polaron is formed as a bound state
within the superconducting gap. 
Unlike the energy of the Majorana fermions, the binding energy of the
polarons is not zero (even in the thrmodynamic limit). 

In Fig.\,\ref{fig:polaron}, we show the density profile, highlighting the occurrence of two hole polarons (upper panel), as well as the polaron profile (lower panel), fitted by the strong-coupling expression: \cite{squeezing}
\begin{equation}
n_\mathrm{polaron}(x)=\frac{1}{2\ell}\,\mathrm{sech}^2\left(\frac{x-x_0}{\ell}\right).
\end{equation}
The area under this profile is equal to 1, showing indeed that the polaron traps one hole, $x_0$ is the location of the polaron, and $\ell$ is the spatial extension of the polaron. The fitting value in the lower panel of Fig.\,\ref{fig:polaron} is $\ell\approx 2.76$ for both polarons.

Outside the coexistence region, the polarons disappear, showing that their existence is really promoted by the interfaces between two coalescing droplets of the high-density phase in the coexistence region, whenever one single hole is trapped at their interface.

\vspace{1cm}
\acknowledgments{The authors are indebted to M. Grilli, N. Scopigno, and M. V. Mazziotti, for their contributions at the early stages of this work. S. Caprara acknowledges financial support from the University of Rome Sapienza, under the Ateneo Projects RM123188E830D258,
RM124190C54BE48D, and RP125199B9FDBFE4. S. Ciuchi acknowledges the hospitality of Rome Sapienza and the funding from  NextGenerationEU National Innovation Ecosystem grant ECS00000041 - VITALITY - CUP E13C22001060006 and grant PE00000023 - IEXSMA - CUP E63C22002180006}
\bibliography{biblio}
\end{document}